\documentclass[11pt]{article}
\usepackage[dvips]{color}
\usepackage{epsfig}
\usepackage{amsmath}
\usepackage{graphicx}

\begin{document}
\title{Entropic force law in the presence of a noncommutative inspired spacetime for a solar system scale}
\author{{S. Hamid Mehdipour$^{a}$\thanks{Email:
mehdipour@liau.ac.ir}}\\
$^{a}${\small {\em  Department of Physics, College of Basic Sciences, Lahijan Branch, Islamic Azad University, }}\\
{\small {\em P. O. Box 1616, Lahijan, Iran.}}} \maketitle

\begin{abstract}
We first study some aspects of a physically inspired kind of a
noncommutative spherically symmetric spacetime based on the
Gaussian-smeared mass distribution for a solar system scale. This
leads to the elimination of a singularity appeared in the origin of
the spacetime. Afterwards, we investigate some features of
Verlinde's scenario in the presence of the mentioned spacetime and
derive several quantities such as Unruh-Verlinde temperature, the
energy and the entropic force on three different types of
holographic screens, namely the static, the stretched horizon and
the accelerating surface.
\end{abstract}

\section{\label{sec:1}Introduction}
It is obvious that there is a connection between gravity and
thermodynamics. In 1995, Jacobson showed that the Einstein field
equations of general relativity are derived from the first law of
thermodynamics \cite{jac}. In 2009, Padmanabhan employed the
equipartition law of energy and the holographic principle to provide
a thermodynamic interpretation of gravity \cite{pad} (to review new
insights into thermodynamical aspects of gravity see \cite{pad2} and
the references therein). Soon after, Verlinde proposed a new idea to
explain the gravity as an entropic force caused by changes in the
information associated with the position of massive particles
\cite{ver}. Verlinde's conjecture has extensively discussed in
various theoretical frameworks such as loop quantum gravity
\cite{smo}, modified gravity \cite{zha}, black hole physics
\cite{tia}, noncommutative geometries \cite{nic}, cosmological
setups \cite{cai}, braneworld scenarios \cite{lin}, Friedmann's
equations with noncommutativity corrections \cite{vfn} and other
fields \cite{oth}.

It is widely believed that the entropy incorporates the emergent
view of gravity with the fundamental microstructure of a quantum
spacetime. Hence, if the origin of gravity is an entropic force, it
is required to take into account the microscopic scale effects by
using accurate tools such as noncommutative geometry (NCG) to
illustrate the microscopic structure of a quantum spacetime, in
Verlinde's scenario. The NCG inspired metrics are a class of
solutions of Einstein equations which include influences of quantum
gravity in very short distances \cite{nic1}. An easy way to explain
this inspired type of NCG theory is to evaluate the mean position of
an object by averaging coordinate operators on appropriate
coordinate coherent states \cite{sma}. In this approach, it has been
shown that the mean position of a point-like object in a
noncommutative manifold is no longer modeled by a Dirac-delta
function distribution, but will be displayed by a Gaussian
distribution of minimal width $\sqrt{\theta}$, where $\theta$ is the
smallest fundamental unit of an observable area in the
noncommutative coordinates, beyond which coordinate resolution is
obscure. Indeed, the emergence of extreme energies at very short
distances in a noncommutative manifold yields the strong quantum
fluctuations which stop any measurements to observe a particle
position with a precision more than an inherent length scale. This
leads to the regular behavior at the origin of black holes. As a
result, a minimal length induced by averaging noncommutative
coordinate fluctuations is appeared in this well-behaved theory of
NCG. In other words, the small scale behavior of point-like
structures is cured such that the particle mass $M$, instead of
being totally localized at a point, is distributed throughout a
region of linear size $\sqrt{\theta}$ as a smeared-like particle (to
study some features of black hole thermodynamics by using the NCG
inspired model, see \cite{nic2}).

On the other hand, the failure of general relativity at the scales
such as solar system has evidently been observed \cite{and}, which
indicates the appearance of dark matter to make compatible
observational probes with the theory \cite{nie}. The important
problem of dark energy and dark matter is connected to their nature
because they behave like ad hoc gravity sources but in a
well-behaved theory of gravity. To avoid this situation and in order
to study the entropic force, we utilize a line element derived from
$f(R)$ gravity theory {\footnote{We assume that a picture of
equilibrium thermodynamics does exist for $f(R)$ gravity theory in
like manner to that in general relativity \cite{bam}.}} for the
solar system scale \cite{saf} (see also \cite{grum}). We include the
noncommutativity correction in the metric derived in \cite{saf} and
find the entropic force. To this purpose, we use three various kinds
of holographic screens, placing at equipotential surfaces, of a
noncommutative inspired spacetime for a solar system scale: the
static holographic screen, stretched horizon and the accelerating
surface. Here it should be noted that for a static 4D metric, the
generic results were obtained by Myung and Kim in Ref.~\cite{myun}
and also by Sakalli in Ref.~\cite{saka}. However, the aim of this
paper is to gain insights into the noncommutative inspired spacetime
in the solar system scale. Throughout the paper, natural units are
used, i.e. $\hbar = c = G = k_B = 1$, and Greek indices run from 0
to 3.

\section{\label{sec:2}Noncommutative inspired spacetime for the solar system}
As a fundamental picture of a quantum spacetime, the
noncommutativity is applied to exhibit the fuzziness of the
spacetime via the commutation relation
\begin{equation}
\label{mat:1} [\textrm{x}^A , \textrm{x}^B]  = i \theta^{AB},
\end{equation}
with a parameter $\theta$ which measures the amount of the
coordinate noncommutativity in the coordinate coherent states
approach \cite{sma}. In the simplest case, $\theta^{AB}$ is an
anti-symmetric, real, $D \times D$ matrix ($D$ is the dimension of
the spacetime). The physical interpretation of $\theta^{AB}$ is the
smallest fundamental unit of an observable area in the $AB$-plane,
in the same way as Planck constant $\hbar$ interprets the smallest
fundamental unit of an observable phase space in quantum mechanics.
Consequently, the resulting geometry is pointless and the concept of
the point is no longer meaningful.

The method we choose here is to look for a static, asymptotically
flat, spherically symmetric, minimal width, Gaussian distribution of
mass whose noncommutative size is governed by the parameter
$\sqrt{\theta}$. For this purpose, we choose the mass density as a
smeared delta-function
\begin{equation}
\label{mat:2}\rho_{\theta}(r)=\frac{M}{(4\pi\theta)
^{\frac{3}{2}}}\exp\bigg(-\frac{r^2}{4\theta}\bigg).
\end{equation}
This result is indeed owing to the fundamental uncertainty encoded
in the coordinate commutator (\ref{mat:1}). The smeared mass
distribution $M_{\theta}$ is found to lead to the result
\begin{equation}
\label{mat:3}M_\theta=\int_0^r\rho_{\theta}(r)4\pi
r^2dr=M\left[{\cal{E}}\left(\frac{r}{2\sqrt{\theta}}\right) -\frac{
r}{\sqrt{\pi\theta}}e^{-\frac{r^2}{4\theta}}\right],
\end{equation}
where ${\cal{E}}(n)$ is the Gaussian error function defined as $
{\cal{E}}(n)\equiv 2/\sqrt{\pi}\int_{0}^{n}e^{-x^2}dx$. In the
regime that noncommutative fluctuations are insignificant,
$r/\sqrt{\theta}\rightarrow\infty$, the Gaussian error function is
equal to one and one recovers the ordinary mass totally localized at
a point, i.e. $M_\theta\rightarrow M$. This means that if
$\sqrt{\theta}$ is too small, the background geometry is described
as a smooth differential manifold and the smeared-like mass descends
to the point-like mass. However, in the limit
$r\rightarrow\sqrt{\theta}$, the metric deviates predominantly from
the standard one and provides new physics at the small scale.

Saffari and Rahvar \cite{saf} solved the field equation in the
vacuum for the spherically symmetric metric and obtained the
dynamics in the solar system and the galactic scales. The line
element for the solar system scale up to the first order in $\alpha$
and under the condition $r\ll d$ is given by
\begin{equation}
\label{mat:4}ds^2=-\left(1-\frac{2M}{r}+\frac{\alpha}{d}r\right)dt^2+
\left(1-\frac{2M}{r}+\frac{\alpha}{d}r\right)^{-1}dr^2+r^2
d\Omega^2,
\end{equation}
where $d\Omega^2$ is the line element on the 2-dimensional unit
sphere. The parameter $\alpha$ is a small dimensionless constant and
$d$ is a characteristic length scale in the order of galactic size.
Here, it should be emphasized that the metric above is extracted
from an ansatz proposed by the authors in Ref.~\cite{saf} for the
derivative of action as a function of distance from the center
\begin{equation}
\label{mat:4.1}F(r)=\left(1+\frac{r}{d}\right)^{-\alpha}.
\end{equation}
It is clear that the case of $\alpha = 0$ yields the standard
Einstein-Hilbert action and therefore the Schwarzschild metric is
recovered. As already pointed out above, the condition for the
existence of the extra term $(\alpha/d)r$ in the metric
(\ref{mat:4}) is $r\ll d$. This means that if $r\gg d$ or
$r\rightarrow\infty$, then the extra term does not exist and this
leads to an asymptotically flat spacetime.

\begin{figure}[htp]
\begin{center}
\includegraphics{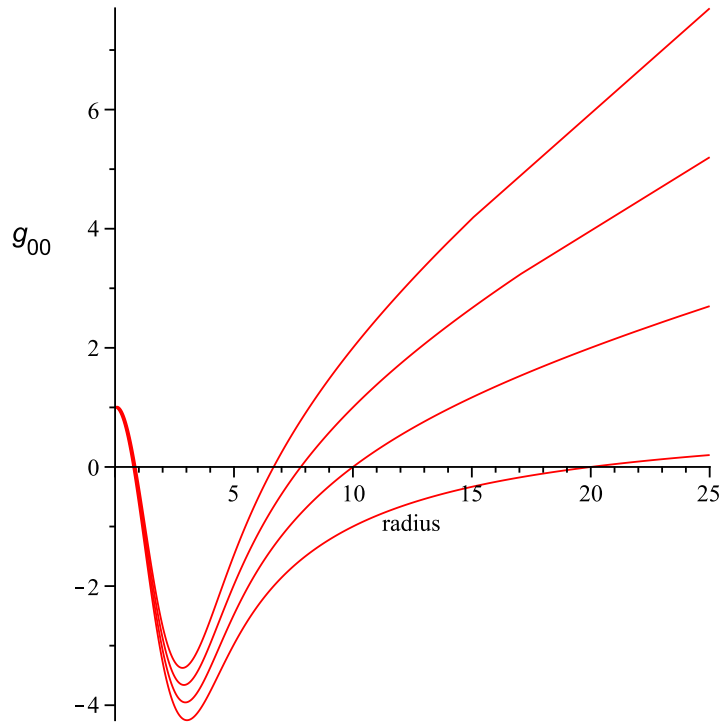}
\end{center}
\vspace{9.6 cm} \caption{\scriptsize {$g_{00}$ versus the radius
$r/\sqrt{\theta}$, for different values of $\alpha/d$. We have set
$M=10.0\sqrt{\theta}$. On the right-hand side of the figure, curves
are marked from bottom to top by $\alpha/d = 0,~
0.10/\sqrt{\theta},~ 0.20/\sqrt{\theta},$ and $0.30/\sqrt{\theta}$.
This figure shows that, the singularity at the origin is eliminated
due to the noncommutativity.}} \label{fig:1}
\end{figure}

Since the extra term $(\alpha/d)r$ is extremely small, therefore in
order to find the noncommutative inspired line element associated
with smeared mass sources, with a good approximation, one can simply
plug the explicit form for the smeared mass distribution into the
metric (\ref{mat:4}) as follows:
\begin{equation}
\label{mat:5}ds^2=-\left(1-\frac{2M_\theta}{r}+\frac{\alpha}{d}r\right)dt^2+
\left(1-\frac{2M_\theta}{r}+\frac{\alpha}{d}r\right)^{-1}dr^2+r^2
d\Omega^2.
\end{equation}
The line element (\ref{mat:5}) characterizes the geometry of a
noncommutative spherically symmetric spacetime for the solar system
scale. Since $\alpha/d$ is very small, the above metric displays a
similar behavior to the noncommutative Schwarzschild (NCS) spacetime
\cite{nic1}. Depending on the different values of mass $M$, the
metric displays three possible causal structures. We plot the
possibility of having two distinct horizons when the mass is large
enough (or $M>M_0$, where $M_0$ is the minimal nonzero mass
corresponding to the minimal nonzero radius $r_0/\sqrt{\theta}$
{\footnote{The other possibilities, i.e. $M\leq M_0$, provide less
good agreement with the usual physical sense \cite{meh}.}}). This
possibility is shown in Fig.~\ref{fig:1}. As the figure shows, the
existence of a minimal nonzero mass and the disappearance of a
divergence at the origin are simply seen in the same manner as the
NCS black hole. There are two horizons, an inner (or noncommutative)
$r_i$ and an outer horizon $r_o$. In the limit
$r/\sqrt{\theta}\rightarrow\infty$, and for $\alpha=0$, the inner
horizon tends to zero, while the outer horizon approaches the
Schwarzschild value, $r_o \rightarrow 2M$. It is clear that if
$\alpha/d$ is too small, then the curves of $g_{00}$ in the NCS
metric (with $\alpha=0$) and the metric (\ref{mat:5}) are extremely
close to each other. As can be seen from the figure, when $\alpha/d$
changes the minus peak of the curves corresponding to $r_0$ is
almost fixed. However, the distance between horizons decreases with
increasing the values of $\alpha/d$. In other words, as $\alpha/d$
deviates from the zero, the outer horizon decreases but the inner
horizon and the minimal nonzero radius remain nearly intact. It is
important to note that, for short distances or high energies there
is a crucial deviation from the standard Schwarzschild metric. In
the next section we use the metric (\ref{mat:5}) to investigate the
entropic picture of gravity in the spacetime under study.

\section{\label{sec:3}Static holographic screen and the entropic force}
In order to obtain the entropic force, we should find the timelike
Killing vector of the metric (\ref{mat:5}). Using the Killing
equation
\begin{equation}
\label{mat:6}\partial_\mu\xi_\nu+\partial_\nu\xi_\mu-2\Gamma^\lambda_{\mu\nu}\xi_\lambda=0,
\end{equation}
with the condition of static spherically symmetric
$\partial_0\xi_\mu=\partial_3\xi_\mu=0$, and also the infinity
condition $\xi_\mu\xi^\mu=-1$, the timelike Killing vector is
written as
\begin{equation}
\label{mat:7}\xi_\mu=\left(\frac{2M_\theta}{r}-\frac{\alpha}{d}r-1,\,0,\,0,\,0\right).
\end{equation}
To define a foliation of space, and distinguishing the holographic
screens $\Omega$ at surfaces of constant redshift, we write the
generalized Newtonian potential $\phi$ in the general relativistic
framework
\begin{equation}
\label{mat:8}\phi=\frac{1}{2}\log\left(-g^{\mu\nu}\xi_\mu\xi_\nu\right),
\end{equation}
where $e^\phi$ is the redshift factor and is equal to one at the
reference point with $\phi = 0$ at infinity. Hence, the acceleration
$a^\mu$ is given by
\begin{equation}
\label{mat:9}a^\mu=-g^{\mu\nu}\nabla_\nu\phi=\left(0,\,2\pi
T,\,0,\,0\right).
\end{equation}
The temperature $T$ on the holographic screen is given by
Unruh-Verlinde temperature connected to the proper acceleration of a
particle close to the screen which can be written as \cite{ver}
\begin{equation}
\label{mat:10}T=-\frac{1}{2\pi}e^\phi n^\mu
a_\mu=\frac{e^\phi}{2\pi}\sqrt{g^{\mu\nu}\nabla_\mu\phi\nabla_\nu\phi},
\end{equation}
where
$n^\mu=\nabla^\mu\phi/\sqrt{g^{\mu\nu}\nabla_\mu\phi\nabla_\nu\phi}$
is a unit vector which is normal to the holographic screen and to
$\xi_\mu$. The Unruh-Verlinde temperature for the metric
(\ref{mat:5}) is simply achieved and reads
\begin{equation}
\label{mat:11}T=\frac{1}{2\pi}\left(\frac{M_\theta}{r^2}-\frac{Mr}{2\sqrt{\pi\theta^3}}e^{-\frac{r^2}{4\theta}}
+\frac{\alpha}{2d}\right).
\end{equation}
The energy on the holographic screen $\Omega$ for a spherically
symmetric spacetime becomes
\begin{equation}
\label{mat:12}E=\frac{1}{4\pi}\int_\Omega e^\phi\nabla\phi dA,
\end{equation}
where $A=4\pi r^2$ is the area of the screen. Hence, the
equipartition of energy rule is given by
\begin{equation}
\label{mat:12.1}E=\frac{1}{2}AT=2ST,
\end{equation}
where $S = A/4$ is the entropy of the holographic screen. Using
Eq.~(\ref{mat:11}), the energy on the screen then takes the form
\begin{equation}
\label{mat:13}E=M_\theta-\frac{Mr^3}{2\sqrt{\pi\theta^3}}e^{-\frac{r^2}{4\theta}}+\frac{\alpha}{2d}r^2.
\end{equation}
The entropic force on the static holographic screen is found to be
\begin{equation}
\label{mat:14}F_\mu=T\nabla_\mu S,
\end{equation}
where $\nabla_\mu S=-2\pi m n_\mu$ is the change in entropy for the
test mass $m$ at a fixed place close to the screen. Finally, the
entropic force in the presence of the noncommutative spherically
symmetric metric for the solar system scale has the form
\begin{equation}
\label{mat:15}F=\sqrt{g^{\mu\nu}F_\mu F_\nu}=
\frac{mM_\theta}{r^2}-\frac{mMr}{2\sqrt{\pi\theta^3}}e^{-\frac{r^2}{4\theta}}+\frac{\alpha
m}{2d}.
\end{equation}
The above result is compatible with the result of Ref.~\cite{saf}.
However, our result is cured in the limit $r\rightarrow 0$. The
corresponding relation for the force in Refs.~\cite{saf,grum} shows
a divergence which appears in the origin of the spacetime under
study. Utilizing the noncommutativity of coordinates, we eliminate
this kind of singularity from the model. The second term at the
right hand side of Eq.~(\ref{mat:15}) is a constant force which is
independent of the source mass $M$. In the limit
$\theta\rightarrow0$, and for $\alpha=0$, the standard results for
the acceleration, the local temperature, the energy, and the
entropic force on the static holographic screen are recovered,
respectively, as follows \cite{liu}:
\begin{equation}
\label{mat:16}a^\mu=\left(0,\,\frac{M}{r^2},\,0,\,0\right),\qquad
T=\frac{M}{2\pi r^2},\qquad E=M,
\end{equation}
and
\begin{equation}
\label{mat:17}F=\frac{mM}{r^2},
\end{equation}
which is the correct Newtonian force for the Schwarzschild metric.

\begin{figure}[htp]
\begin{center}
\includegraphics{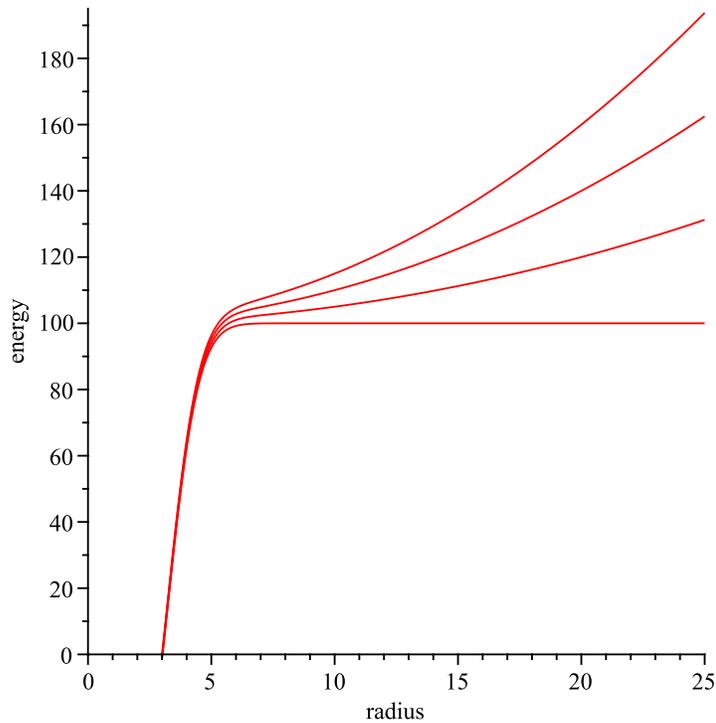}
\end{center}
\vspace{9.6 cm} \caption{\scriptsize {The energy, $E/\sqrt{\theta}$,
versus the radius, $r/\sqrt{\theta}$, for different values of
$\alpha/d$. We have set $m=1.0\sqrt{\theta}$ and
$M=100.0\sqrt{\theta}$. On the right-hand side of the figure, curves
are marked from bottom to top by $\alpha/d = 0,~
0.10/\sqrt{\theta},~ 0.20/\sqrt{\theta},$ and
$0.30/\sqrt{\theta}$.}} \label{fig:2}
\end{figure}

\begin{figure}[htp]
\begin{center}
\includegraphics{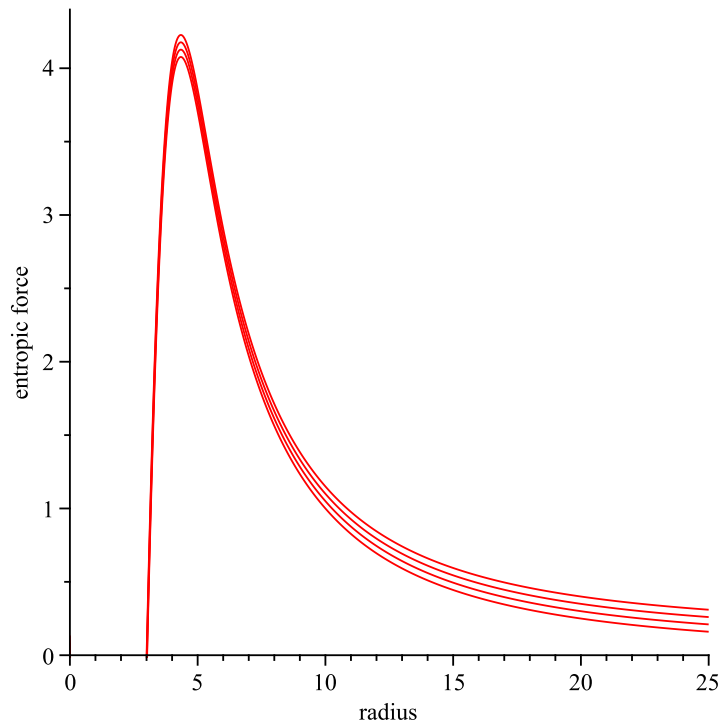}
\end{center}
\vspace{9 cm} \caption{\scriptsize {The entropic force $F$ versus
the radius, $r/\sqrt{\theta}$, for different values of $\alpha/d$.
We have set $m=1.0\sqrt{\theta}$ and $M=100.0\sqrt{\theta}$. On the
right-hand side of the figure, curves are marked from bottom to top
by $\alpha/d = 0,~ 0.10/\sqrt{\theta},~ 0.20/\sqrt{\theta},$ and
$0.30/\sqrt{\theta}$.}} \label{fig:3}
\end{figure}

The numerical computation of the energy and the entropic force as a
function of the radius for some different values of $\alpha/d$ are
depicted in Figs.~\ref{fig:2} and \ref{fig:3}, respectively.
Fig.~\ref{fig:2} shows the energy increases with deviating the
coefficient $\alpha/d$ from the zero. Similarly, as can be seen from
Fig.~\ref{fig:3}, as the coefficient $\alpha/d$ deviates from the
zero, the entropic force increases but the peak in the entropic
force in the vicinity of the minimal non-zero radius $r_0$ remains
nearly intact. It is worth noting that, these numerical results are
fairly similar to the results reported in our previous work
\cite{mehdi}. In Ref.~\cite{mehdi}, the thermodynamical features of
a non-commutative inspired Schwarzschild-anti-deSitter black hole in
the context of entropic gravity model have been studied. It seems
the behavior of the negative cosmological constant and the
coefficient $\alpha/d$ associated with the Pioneer anomaly in the
solar system scale are the same.

It must be stressed here that the case of $r<r_0$ leads to some out
of the standard dynamical features like negative entropic force,
i.e. gravitational repulsive force, and negative energy; as a
result, one should make the requirement that $E\geq 0$ \cite{meh}.
Therefore, the appearance of a lower finite cut-off at the
short-scale gravity compels a bound on any measurements to determine
a particle position in a non-commutative gravity theory.

\section{\label{sec:4} Accelerating surface and the entropic force}
In this section, we are interested in introducing the entropic force
on the accelerating surface of a noncommutative inspired spacetime
for a solar system scale. The accelerating surface was first
introduced in \cite{make}. Here, the accelerating surface plays the
role of the holographic screen and we consider the accelerating
surface as the accelerating screen. By defining a future pointing
unit vector $u^\mu$, which is the congruence for the timelike world
lines of the points on a spacelike hypersurface $S^2$, the
orthogonality condition
\begin{equation}
\label{mat:18}u^\mu n_\mu=0,
\end{equation}
should be satisfied, where $n_\mu=[0,1/\sqrt{-g_{00}},0,0]$ is the
normal vector on $S^2$. Using the vector $u^\mu$, one can find the
change of the heat, which is associated with a proper acceleration
vector field as follows:
\begin{equation}
\label{mat:19}a^\mu=u^\nu \nabla_\nu u^\mu,
\end{equation}
The only nonzero component of the future pointing unit vector is
\begin{equation}
\label{mat:20}a^1=u^0 \nabla_0 u^1.
\end{equation}
The proper acceleration is found to be
\begin{equation}
\label{mat:21}a=a^\mu
n_\mu=\frac{M_\theta}{r^2}-\frac{Mr}{2\sqrt{\pi\theta^3}}e^{-\frac{r^2}{4\theta}}+\frac{\alpha}{2d}.
\end{equation}
The Unruh temperature or so-called bulk temperature on the
accelerating screen is defined by
\begin{equation}
\label{mat:22}T=\frac{a}{2\pi},
\end{equation}
which means that an accelerating observer on the accelerating screen
observes thermal radiation with the Unruh temperature. It is evident
that the screen temperature is identical to the bulk temperature and
thus its corresponding entropic force on a test mass $m$ in the
vicinity of the accelerating screen is equal to Eq.~(\ref{mat:15}).

\section{\label{sec:5} Stretched horizon and the entropic force}
In this section, we consider the stretched horizon to be a
holographic screen. In this way, the entropic force on the stretched
horizon will be obtained. The idea of stretched horizon \cite{suss}
is interesting due to its location. Its location is extremely
adjacent to the event horizon. In other words, all thermodynamical
quantities are measured by an observer existing at the proper
distance $l_{pl}$ away from the horizon. The radial distance of the
stretched horizon from the center of the spacetime under study is
determined as $r = r_o + 1/r_o$ such that $r_o \gg1$. The near
horizon limit and the geometry of a stretched horizon can be
characterized by a Rindler spacetime. Hence, the local Rindler
temperature is immediately written by
\begin{equation}
\label{mat:23}T=\frac{1}{4\pi}\frac{1}{\sqrt{-g_{00}}}\bigg|\frac{dg_{00}}{dr}\bigg|\Bigg|_{r=r_o+\frac{1}{r_o}}.
\end{equation}
The length contraction of $1/r_o = \sqrt{-g_{00}}$ is as a result of
the redshift transformation close to the horizon. Thus, the local
temperature on the stretched horizon takes the following form
\begin{eqnarray}
\label{mat:24}T=\frac{1}{2\pi}\left(\frac{M_\theta(r=r_o+1/r_o)}{(r_o+1/r_o)^2}+\frac{\alpha}{2d}\right)
-\frac{M(r_o+1/r_o)}{2\sqrt{\pi\theta^3}}e^{-\frac{(r_o+1/r_o)^2}{4\theta}}
\nonumber
\\ \times\left(1-\frac{2M_\theta(r=r_o+1/r_o)}{r_o+1/r_o}+\frac{\alpha}{d}(r_o+1/r_o)\right)^{-\frac{1}{2}}.
\end{eqnarray}
It is important to note that all temperatures on the stretched
horizon acquired via different methods are the same in the leading
order. So, we have
\begin{equation}
\label{mat:25}T=\frac{M
e^{-\frac{r_o^2}{4\theta}}}{4(\pi\theta)^{\frac{3}{2}}}r_o.
\end{equation}
In the Rindler space, the local energy yields
\begin{equation}
\label{mat:26}E=\frac{AM
e^{-\frac{r_o^2}{4\theta}}}{8(\pi\theta)^{\frac{3}{2}}}r_o.
\end{equation}
Finally for the entropic force, we obtain
\begin{equation}
\label{mat:27}F=\frac{mM
e^{-\frac{r_o^2}{4\theta}}}{2\sqrt{\pi\theta^3}}r_o=ma,
\end{equation}
where $a=\frac{M
e^{-\frac{r_o^2}{4\theta}}}{2\sqrt{\pi\theta^3}}r_o$ is the proper
acceleration on the stretched horizon. Among three distinct screens
we have introduced the stretched one is special due to the fact that
it is located at a special location and it is clear that when all
the other entropic forces are recomputed for the place of the
stretched horizon they all turn into the same \cite{myun,saka}.

\section{\label{sec:4}Summary}
Recently, Verlinde has proposed a new idea of duality between
thermodynamics and gravity which yields an emergent phenomenon for
the origin of gravity. Since the entropy unites the emergent picture
of gravity with the fundamental microstructure of a quantum
spacetime, it is therefore essential to take into account the
microscopic scale effects via exact tools such as the theory of NCG
to clarify the microscopic structure of a quantum spacetime in
Verlinde's proposal. For this purpose, we have used the NCG inspired
model representing smeared structures to incorporate the microscopic
structure of the spacetime with the entropic view of gravity. As a
result, the singularity at the origin of the spacetime is removed.
We have defined three different surfaces, which are individually
such a candidate for the holographic screen of a NCG inspired
spacetime. Those are the static holographic screen, the accelerating
surface, and the stretched horizon or the Rindler spacetime of the
NCG inspired model. In this setup, some thermodynamical quantities
of the noncommutative spacetime, e.g. the entropic force for the
solar system scale have been derived. We have observed that if one
consider the accelerating surface as the accelerating screen, it is
very natural to define the acceleration of a test mass through the
Unruh temperature and therefore an entropic force can be correctly
retrieved. Among these screens, the stretched horizon is particular.
Since, the stretched horizon is situated at such a special location,
therefore when all the other entropic forces are recomputed for that
special place, they all become similar.\\

\section*{Acknowledgments}
The author is indebted to R. Saffari for some inspiring
discussions.\\

\end{document}